\documentclass[12pt,a4paper]{article} 
\usepackage{ae}
\usepackage{graphicx} 
\usepackage{natbib}
\usepackage{setspace} 
\title{Unexpected trends in the social facilitated survival of termites} 
\vspace{4cm}
\author{Og DeSouza 
\\Departamento de Biologia Animal 
\\Universidade Federal de Viçosa 
\\36571-000 Viçosa MG, Brazil
\\isoptera@insecta.ufv.br
\\Fax: +55 31 3899 2537 Tel: +55 31 3899 2532
\vspace{1cm}
\\ \&
\vspace{1cm}
\\ Octavio Miramontes \\Departamento de Sistemas Complejos
\\Instituto de Fisica \\Universidad Nacional Autónoma de México
\\Aptdo. Postal 20-364 \\México 01000 DF, México
\\octavio@fenix.ifisicacu.unam.mx}
\vspace{1cm}

\date{}
\begin{document}
\maketitle

{\bf Running head:} Unexpected trends in termite survival \vspace{1cm}


\newpage \doublespacing

\begin{abstract}
  Survival in groups of termites is known to be socially facilitated
  since individual longevity is increased together with the size of
  the group. Here we report on experimental evidence that survival as
  a function of group size in {\textit Cornitermes cumulans} (Kollar)
  does not increase asysmptotically but it peaks in a group size range
  at which individual longevity is maximal. Strinkly, this same effect
  in other social insects was noted more than fifty years ago, but it
  was regarded as ``aberrant'' since it has been generally assumed
  that individual survival should increase with the size of the group.
  We also provide evidence that individual activity --measured as
  mobility-- is maximal at approximately the same group-size range as
  the observed survival.
\end{abstract}
\newpage
\section{Introduction}
\cite{grasse+c44} reported a number of experiments to check the
effects of group size on the survival of \textit{Apis} sp.,
\textit{Polistes} sp, \textit{Leptotorax} sp., \textit{Formica} sp.
(Insecta: Hymenoptera) and \textit{Reticulitermes} sp.(Insecta:
Isoptera).  In all of them, but in {\textit Polistes}, survival
increased with increasing group size. In \textit{Polistes} they
reported an ``aberrant'' result in which survival in both large and
small groups was smaller than in the middle ones. The authors
atributed these obeservations to bad experimental procedures or to a
very particular physiological trait of the species. The experiment was
hence repeated several times --even by independent observers-- but the
``aberrant'' result was always found. On a survey experiment performed
on the summer 1994, we test grouped individuals of \textit{Syntermes}
sp., \textit{Nasutitermes} sp., and \textit{Cornitermes} sp. (Insecta,
Isoptera). Preliminary results showed the same unexpected behaviour on
the survival of grouped \textit{Cornitermes cumulans} (Kollar).

Such a convergence of results allow us to suspect that Grassé \&
Chauvin (1944) have in fact unwarrantly spotted a pattern that is
biologically sound. We base our suspicion on previous experimental
works which show that (i) individual survivorship of termites, while
affected by group size, may be related to individual mobility of group
components \citep{miramontes+d96}, and (ii) longevity of individual
termites seems to obey a hump-shaped curve, clearly dependent on some
(yet) unknown attribute of group size \citep{desouza+msb01}. In
addition, there is theoretical evidence that cooperation could either
stabilize or destabilize the dynamics of a social group, leading to
maximal individual fitness at intermediate colony sizes
\citep{aviles99}.

This paper aims to present experimental evidence linking individual
mobility, survivorship, and group size in termites. To achieve this we
show that both individual mobility and survival are humped functions
peaking at similiar group-size range. We argue, on experimental and
theoretical grounds, that survival of starving grouped termite workers
--at least in \textit{Cornitermes} sp-- obeys a humped-shape curve
which present a biologically sound, not fortuitous, relationship to
individual mobility.

\section{Material \& Methods}
To confirm the trend observed in survey experiments, two further
experiments were performed in 1996 and 1997 with grouped termites of
this species collected from wild colonies in Viçosa in the state of
Minas Gerais, Southeastern Brazil. The first experiment, spanning
several days, was designed to measure longevity while the second was
intended to measure individual activity. Both experiments involved
termite workers (third instar and beyond) of different group sizes.

In the longevity experiment, termites from a wild colony were randomly
placed in groups of 1, 2, 4, 8, 12, 16, 20, and 24 (each group
replicated eight times). Collection took place in 04 April 1996. The
groups were confined in test tubes made of transparent glass (9.5
$\times$ 1.4cm) with hermetically sealing rubber caps. Each test tube
is a `replicate'. Test tubes have been previously washed and rinsed,
soaked in sodium hypochlorin for 24h, rinsed again, and sterilised at
$180\,^{\circ}\mathrm{C}$ for one hour. Tubes with termites were kept
horizontally, separated by plastic foam to prevent stridulation or
other mechanically transmitted signals to propagate between the tubes.
The workers were incubated in the dark at a constant temperature
chamber ($25\,^{\circ}\mathrm{C}\pm0.5$) and were allowed to
acclimatise for 12h. No food or water was provided, but tubes were
opened each 24h, to allow air exchange. Tubes were exposed to light
during the counting of survivors only (no more than 5 min).
Observations were made each eight hours intervals, until all
individuals were dead (153h; 20 observations). Termite groups in which
dead individuals presented any sign of cannibalism were excluded from
data analysis.

In the activity experiment, termites from another wild colony were
randomly placed in groups of 2, 4, 6, 8, 16, 20, and 30 individuals
(each group replicated four times). Collection took place in 12
October 1997. The experiment was setup as above, using a controlled
temperature room ($24\,^{\circ}\mathrm{C}\pm0.5$) rather than a
chamber. Termites were allowed to acclimatise in such a room, for
three hours before the observations began. Two observers took notes of
the number of termites moving in each group, at five minutes
intervals, during ten continuous hours. Each observer managed two
replicates of the experiment (two tubes of each group size). Light
intensity in the room was the lowest possible to allow the work of the
observers.

Statistical analyses aimed to check whether group size would affect
(i) survival and (ii) mobility of termites. Either quantities were
treated as the $y$-variable in separate regression analyses, always
with group size as the $x$-variable. Survival was measured as the
average number of hours spent until individual termites die in a given
group, which was calculated for each replicate (a given test tube
containing termites), using Weibull frequency distribution
\citep{crawley93,pinderIII+ws78}. Survival functions obey the general
form:
\begin{equation}
\ln S(t)=\mu^{-\alpha} t^{\alpha}
\end{equation} 
where $S(t)$ is the proportion of individuals from the initial cohort
(the initial group of termites) still alive at time $t$, $\alpha$ is
the shape parameter, $\mu$ is the mean number of hours until death by
the termites from a given replicate. The first step in the survival
analysis was to establish the value of the shape parameter $\alpha$,
thereby defining whether or not the proportion of individuals dying
was constant through time ($\alpha=1$, $dN/dt=1/\mu$, exponential
fit). To do so, survival curves were tentatively fit to the data,
simulating different shape parameters, one at a time, until the error
deviance reached an asymptote. The model obtained was then compared
with a model assuming $\alpha=1$ (exponential fit), on the basis of
their contribution to the error deviance. Parsimony requires to choose
the exponential fit (simplest model) if it does not differ
significantly ($P$ > 0.05; $\chi^2$ tables; 1 df) from an alternative,
more complete, model.

After defining the shape parameter, the next step was to calculate the
mean number of hours termites spent to die when confined in each group
size ($\mu$ in equation 1 above). Each of such means represent the
number of hours spent to die averaged across all individuals belonging
to a given replicate, and are referred to as ``mean time to death''.
Mean times to death for each replicate were then collapsed into a
single arithimetic mean for each group size, a valid procedure to
avoid pseudoreplication effects \citep{crawley93}. Such a procedure
produces eight means (each one represented by a dot in Fig.1), one for
each group size.

Means thereby produced were used to check whether termite survival
(y-var) would be affected by group size (x-var). To do so, models were
fitted to the data, starting from a null model ($y=b$, were $b$ is the
grand mean), and adding new terms until achieving the best trade-off
between percent of variance explained ($r^2$), and $P$-values obtained
(see Table ~\ref{tab:survival}).

The general pattern of mobility of termites in a given replicate (a
test tube containing termites) was measured by fitting a simple
regression line of the form $y=b+ax$, through the data points formed
by plotting the number of moving termites (y-var) against time in
minutes (x-var). The slope $a$ of each line was used as a measurement
of the general pattern of mobility for termites belonging to that
replicate. If the number of moving termites increases as times goes
by, the curve fitted to the data would show a positive slope.
Conversely, a negative slope would indicate a decay in group mobility.
When no clear trend is to be observed for a given replicate, its
``mobility curve'' would show a null slope. For each group size we
obtained four slopes, since there were four replicates (i.e., four
test tubes containing termites). The slope values thereby obtained
were collapsed into a single arithimetic mean for each group size
aiming, as above, to avoid pseudoreplication effects. Such a procedure
produces seven means (each one represented by a dot in Fig.2), one for
each group size. Such means are refereed to as the ``general pattern
of mobility'' for that group size.

Means thereby produced were used to check whether the general pattern
of mobility (slope $a$ of mobility lines, see above) would depend on
the size of the group were termites have been confined. Models were
fitted and selected as above (see Table \ref{tab:mobility}).

All statistical analyses were performed using R \citep{ihaka+g96}.

\section{Results}
The mean time to death of starved termites obeys a humped function of
group size, presenting a peak at a characteristic group size and
decreasing at both smaller and larger densities (Polynomial model,
order four: F[4;3] = 20.22 ; $r{^2}adj$ = 0.916; $P$ = 0.02; Fig.  1;
Table ~\ref{tab:survival}) Accordingly, group activity presents a
similar dependency on group size, showing a peak at an intermediate
density and decreasing at both smaller and larger values (Quadratic
model: F[2;4] = 9.095; $r{^2}adj$ = 0.730; $P$ = 0.03; Fig. 2; Table
~\ref{tab:mobility}). This quadratic model explains the data better
than the simple linear one: adding a quadratic term to the linear
model, increases by 18.7\% the percentage of variance explained and
improves the $P$-value (Table ~\ref{tab:mobility}).

Such results are consistent across different procedures of data
analysis. As specified above, each dot in Figs.1 and 2 represents an
arithmetic mean taken across the respective replicates. If, however,
we plot the raw data (i.e., splitting the arithmetic means into its
components), we observe the same general shape in both curves (Figs.3
and 4).

\section{Discussion} 
Across the Isoptera, a range of different socially facilitated
behaviours has been extensively reported, although survival is the one
that has attracted more attention (Table \ref{tab:traits}). In all
survival studies, longevity has been reported to increase as group
size increases. This may lead one to intuitively assume an asymptotic
behaviour of such a curve, accepting that very large groups would
attain the maximal possible survival rates. In this sense, the results
reported by Grassé \& Chauvin(1944) in which survival of wasps is
maximal at moderate densities (rather than at large group sizes),
could be considered ``aberrant''. However, a striking similarity can
be observed between such results and those found here for \textit{C.
  cumulans} termites (Fig.  1). More puzziling, termites from
different wild colonies of the same species show a similar humped
function, even when they have been poisoned by insecticide
\citep{desouza+msb01}. Such a convergence of results lead us to
suspect that this pattern is, in fact, biologically sound. It seems
that survival of starving termite workers --at least in
\textit{C.cumulans}-- is not asymptotically related to group size, but
obeys a humped function, in which an optimal density (=number of
individuals per unit area) assures maximal survival.

Optimal densities in social insects have already been predicted as a
consequence of evolutionary pathways maximising individual fittness
\citep{higashi+y93}. Accordingly, extraordinary lifespan is argued to
derive from evolutionary processes in which such insects benefit from
sociality \citep{keller+g77}. Proximate mechanisms through which
optimal densities would lead to better survival, however, are still
poorly understood. Many works dealing with cluster size in group
organisms are based on optimality arguments regarding resource usage,
territory defence, anti-predatory behaviour and disease resistance
\citep{creel97,fritz+g96,giraldeau+b99,roberts96,rosengaus+mct98}.
Others, inspired on novel concepts of the sciences of complex systems,
would invoke a relationship between group-size, information fluxes and
task performance
\citep{adler+g92,bonabeau+td98,deneubourg+gpd86,gordon96,pacala+gg96}.
The interconnected nature of such systems is thought to be the basis
for the self-organisation of a variety of cluster-related phenomen
\citep{bonabeau+td97,camazine91,miramontes+sg93,theraulaz+bc95}.
Specifically for termites, evidences seem to point out that survival
is related to mobility \citep{miramontes+d96}. In addtion, studies
addressing interindividual interactions and self-orgnisation predict
the existence of group-size effects causing information transfer and
behavioural diversity to reach near optimal conditions
\citep{miramontes95,sole+m95}. Accordingly, termites studied here
present maximal values for group activity at intermediate densities
(Fig. 2), which lie in the same group-size range as the observed
survival (Fig.1). It seems,therefore, that the association between
group-size, non-asymptotic survival and greater individual activity is
not fortuitous in \textit{C.cumulans}. Moreover, to the extent that
mobility affects the rate of social contacts --and thereby,
cooperation-- our results seem to agree to those of \cite{aviles99},
who show an explicit link between cooperation, individual fitness, and
group size in social organisms.

These results point out to specific mechanisms present in the dynamics
of groups that may act to regulate observables such as density, number
of individuals and rate of social contacts, that in turn have an
impact on collective task performance. Arguably, behaviours such as
colony fission, swarming, cannibalism, or resting, may be the natural
mechanisms responsible for the regulation of the proportion of
individuals in the colony. In other words, the patterns presented here
may have unexpected consequences for understanding the life cycles of
these organisms, and therefore are worth exploring further.

\section{Acknowledgements}
We thank CA Santos and DL Bernardo for assistance in laboratory and
field work. Dr. A Chopps kindly refreshed our ideas with new and deep
insights. OM would like to thank MP Hassel and P Rohani for assistance
while visiting the UFVi\c cosa and the Termitology Laboratory at UFV.
ODS would like to thank the staff of the Departamento de Sistemas
Complejos (IFisica, UNAM) for all the help during an academic visit.
This work has been supported by UNAM-DGPA \#INI08496, CONACyT
\#3280P-E9607, FAPEMIG, FINEP, and PROIN-Entomologia (CAPES-UFV). This
paper was entirely produced using free software (Linux, \LaTeXe,
Xemacs, and R).
\renewcommand{\refname}{Literature}
\bibliography{/home/og/09Bibliog/bib/bibmaster}
\bibliographystyle{/usr/TeX/texmf/bibtex/bst/inssoc/ins2}

\newpage
\section{Legend to figures}
\begin{description}
\item[Figure 1] Average number of hours termite workers spent to die
  when confined with conspecifics in test tubes, in the absence of
  food. Each dot represents an arithmetic mean across eight replicates
  (eight different test tubes). Average number of hours for each
  replicate was calculated by censored survival analysis, with Weibull
  distribution.
\item[Figure 2] General pattern of mobility of termites confined with
  conspecifics in test tubes, in the absence of food. Mobility
  patterns are characterized here using the slope of a line fitted
  across a scatterplot of the number of moving termites (y-var) versus
  time (x-var), four each of the group sizes tested. Positive slopes
  denote that the number of moving termites increases with time,
  negative slopes denote the opposite. Null slopes denote a constant
  number of moving termites across the time range when observations
  were done. Each dot represents the average of the slopes presented
  by each of four replicates (four different test tubes).
\item[Figure 3] Average number of hours termite workers spent to die
  when confined with conspecifics in test tubes, in the absence of
  food. Each dot represents the average number of hours termites
  survived in a single replicate (a test tube), as calculated by
  censored survival analysis, with Weibull distribution. Within each
  group size, datapoints of this figure were collapsed into a single
  arithmetic mean in order to produce Fig.1.
\item[Figure 4] General pattern of mobility of termites confined with
  conspecifics in test tubes, in the absence of food. Mobility
  patterns are characterized here using the slope of a line fitted
  across a scatterplot of the number of moving termites (y-var) versus
  time (x-var), four each of the group sizes tested. Each dot
  represents the slope presented by a single of four replicates (four
  different test tubes). Within each group size, datapoints of this
  figure were collapsed into a single arithmetic mean in order to
  produce Fig.2.
\end{description}
\newpage
\begin{figure}[!h]
  \centering \includegraphics[width=12cm]{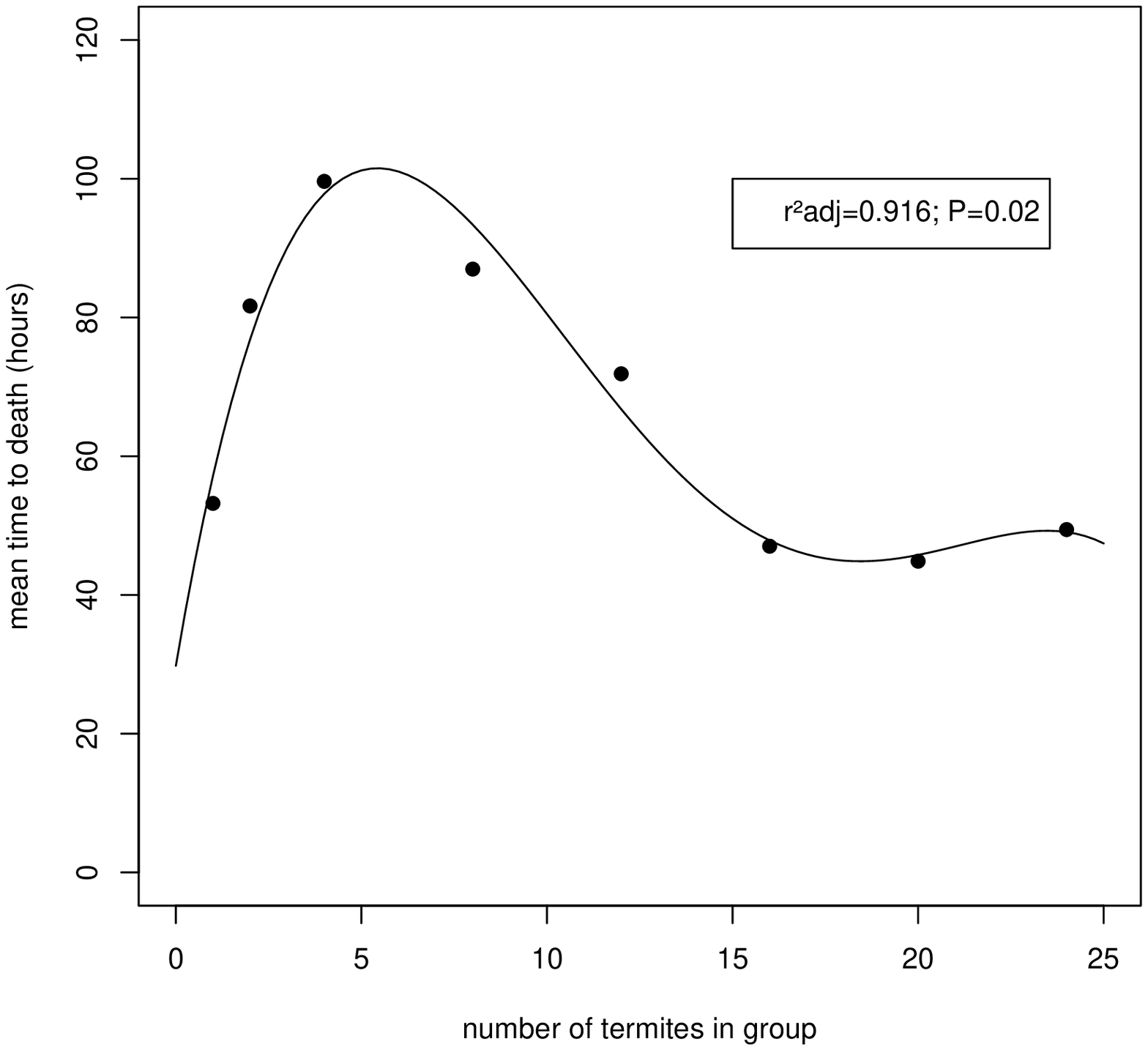}
  \label{fig:surv8}
\end{figure}
\newpage

\begin{figure}[!h]
  \centering \includegraphics[width=12cm]{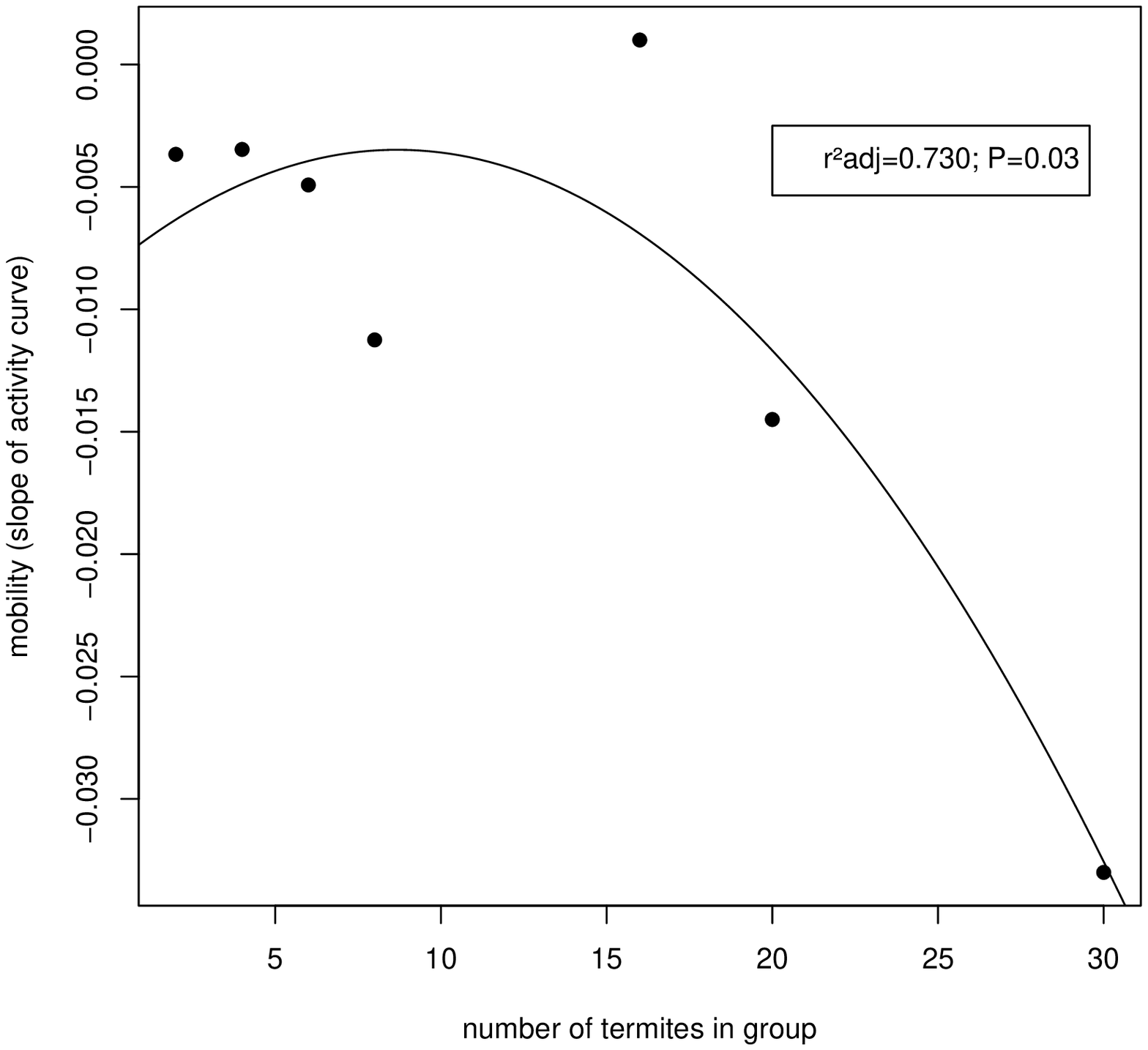}
  \label{fig:mobil7}
\end{figure}

\newpage

\begin{figure}[!h]
  \centering \includegraphics[width=12cm]{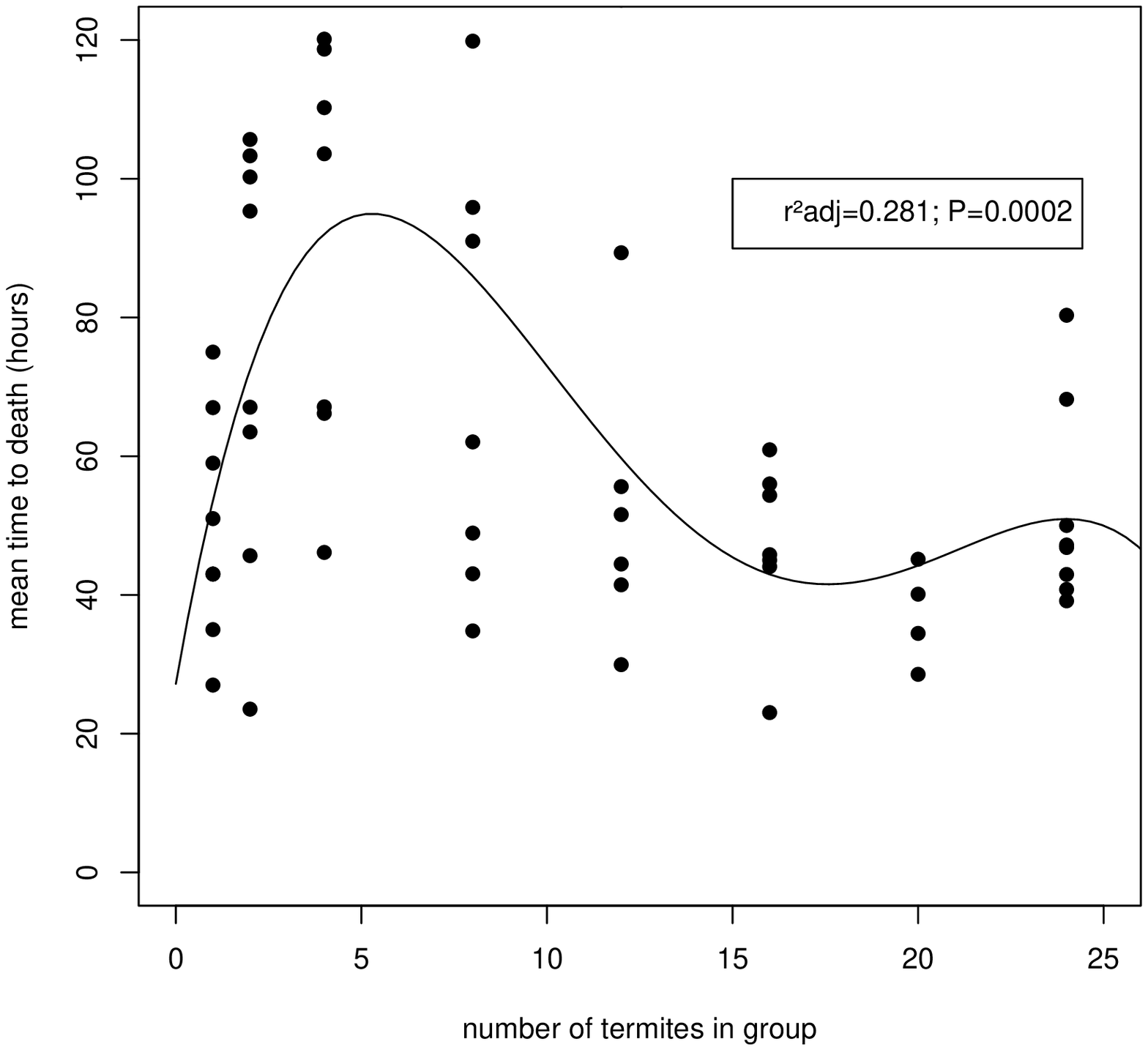}
  \label{fig:surv58}
\end{figure}

\newpage

\begin{figure}[!h]
  \centering \includegraphics[width=12cm]{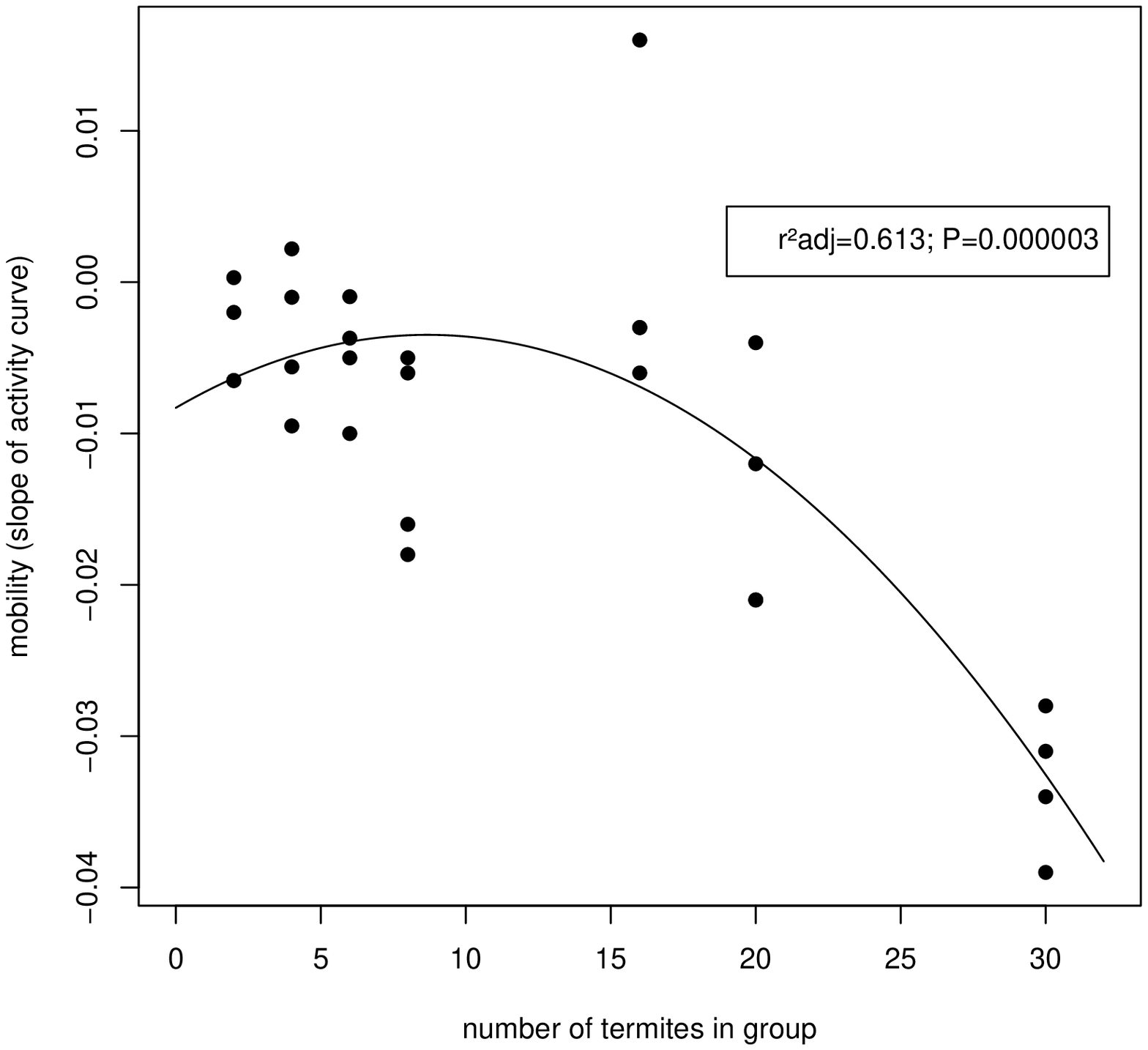}
  \label{fig:mobil28}
\end{figure}

\newpage
\begin{table}[!h] 
  \centering
  \doublespacing
  \caption{Models explaining survival the effect of group size
    ($x$-var) on the mean time to death ($y$-var) of starved termites
    confined in test tubes. Models provided here are intended to be
    illustrative of the general shape of the curves described by the
    data, rather than a statement on its specific poisition when
    plotted. Models were calculated on the datapoints of
    Fig.1. Polynomial model of order 5 is not significant.}
  \begin{tabular}{lccc}
    \hline
    Model                                         &  r$^2$adj  & P\\
    \hline
    $y=66.829$                                    &  -         & 1.00\\
    $y=88.8355-15638x$                            &  0.309     & 0.09\\
    $y=75.8990+0.8722x-0.1016x^2$                  &  0.262     & 0.20\\
    $y=47.94922+16.77077x-1.74097x^2+0.04376x^3$    &  0.798     & 0.02\\
    $y=29.783398+31.566777x-4.431666x^2+0.211764x^3-0.003355x^4$&0.916&0.02\\
    \hline
  \end{tabular}
\label{tab:survival}
\end{table}

\newpage
\begin{table}[!h]
  \centering
  \doublespacing
  \caption{Models explaining survival the effect of group size
    ($x$-var) on the mobility ($y$-var) of starved termites
    confined in test tubes. Models provided here are intended to be
    illustrative of the general shape of the curves described by the
    data, rather than a statement on its specific poisition when
    plotted. Models were calculated on datapoints of
    Fig.2. Polynomial model of order 3 is not significant.}
  \begin{tabular}{lccc}
    \hline
    Model                                         &  r$^2$adj  & P\\
    \hline
    $y=-0.009973$                                    &  -         & 1.00\\
    $y=0.0008778-0.0008832x$                         &  0.543     & 0.04\\
    $y=-8.299 \cdot 10^{-3}+1.110 \cdot 10^{-3}x-6.396 \cdot 10^{-5}x^2$           &  0.730     & 0.03\\
   
    \hline
  \end{tabular}
  \label{tab:mobility}
\end{table}
\newpage
\begin{table}[!h]
  \centering
  \doublespacing
  \caption{Some different behavioural traits that are known to be
    socially facilitated across the Isoptera}
  \begin{tabular}{lll}
    \hline
    Genus   &  Trait   & Reference\\
    \hline
    \textit{Bellicositermes}  &Ovarium development  & \cite{grasse39}\\
    \textit{Bifiditermes} &Food exchange         &\cite{afzal83}\\
    \textit{Cephalotermes}  &Ovarium development  &\cite{grasse39}\\
    \textit{Coptotermes}  &Survival and feeding  &\cite{lenz+w80}\\
    \textit{Cornitermes} &Tolerance to poisoning &\cite{desouza+msb01}\\
    \textit{Cryptotermes} &Survival               &\cite{williams+pj82}\\
    \textit{Kalotermes}  &Nest digging           &\cite{springhetti90}\\
    \textit{Macrotermes}  &Caste differentiation  &\cite{okot-kotber83}\\
    \textit{Nasutitermes} &Survival and feeding   &\cite{lenz+w80}\\
                       &Survival under starvation    &\cite{miramontes+d96}\\
    \textit{Reticulitermes} &Survival               &\cite{grasse+c44}\\
    \textit{Zootermopsis} &Survival under infection   &\cite{rosengaus+mct98}\\
    \hline
  \end{tabular}
\label{tab:traits}
\end{table}

\end{document}